\pdfoutput=1
\documentclass[aps,prl,twocolumn,notitlepage,superscriptaddress,longbibliography,footinbib,preprintnumbers,floatfix]{revtex4-2}
\usepackage{tikz-cd}
\usepackage{tikz}
\usepackage{cancel}
\usetikzlibrary{graphs,quotes}

\usepackage{array}
\newcommand{\runinsec}[1]{\textit{\textcolor{runinred}{#1.---}}\ }

\usepackage{epsf}
\usepackage{times}
\usepackage[colorlinks=true,citecolor=blue,linkcolor=blue, urlcolor=blue]{hyperref} 
\usepackage{slashed}
\usepackage{amsthm}
\usepackage{thmtools}
\usepackage{physics, bm}
\usepackage{mathrsfs}
\usepackage{amsfonts, mathtools, amsmath,amssymb,dsfont}
\usepackage{esvect}
\usepackage{enumerate,multirow,bbold, enumitem}
\usepackage{array}
\usepackage{tikz-cd}
\usepackage{bbding}
\usepackage[normalem]{ulem}
\usepackage{calligra}
\usepackage{bbm}
\usepackage{mathbbol}

\declaretheoremstyle[
shaded={bgcolor=\color{rgb}{0.9,0.9,0.9}}
]{theorem}

\declaretheoremstyle[
shaded={bgcolor=\color{rgb}{0.9,0.9,0.9}}
]{question}

\declaretheoremstyle[
shaded={bgcolor=\color{rgb}{0.9,0.9,0.9}}
]{remark}

\declaretheoremstyle[
shaded={bgcolor=\color{rgb}{0.9,0.9,0.9}}
]{proposition}

\declaretheoremstyle[
shaded={bgcolor=\color{rgb}{0.9,0.9,0.9}}
]{definition}

\declaretheoremstyle[
shaded={bgcolor=\color{rgb}{0.9,0.9,0.9}}
]{assumption}

\declaretheoremstyle[
shaded={bgcolor=\color{rgb}{0.9,0.9,0.9}}
]{conjecture}

\declaretheoremstyle[
shaded={bgcolor=\color{rgb}{0.9,0.9,0.9}}
]{corrorary}

\declaretheoremstyle[
shaded={bgcolor=\color{rgb}{0.9,0.9,0.9}}
]{axiom}

\declaretheoremstyle[
shaded={bgcolor=\color{rgb}{0.9,0.9,0.9}}
]{lemma}

\declaretheoremstyle[
shaded={bgcolor=\color{rgb}{0.9,0.9,0.9}}
]{problem}

\newcommand{\RNum}[1]{\uppercase\expandafter{\romannumeral #1\relax}}

\newcommand{\Rom}[1]{ \uppercase\expandafter{\romannumeral#1}}

\usepackage{graphicx}
\usepackage{orcidlink}
\definecolor{runinred}{rgb}{0.75,0,0}

\newcommand{\xsout}{\bgroup \markoverwith{\hbox{\rule[.5ex]{2pt}{.4pt}\llap{\rule[-.35ex]{2pt}{.4pt}}}}\ULon}

\begin{document}

\title{Mimicry of \texorpdfstring{$d$}{d}-wave Andreev Signatures by Interlayer Pairing}

\author{Zhongyi Zhang\,\orcidlink{0000-0003-0887-9358}}\email{zhongyizhang@ust.hk}
\affiliation{Department of Physics, Hong Kong University of Science and Technology, Clear Water Bay, Hong Kong, China}

\author{Shengshan Qin\,\orcidlink{0009-0003-3484-8747}}
\affiliation{School of Physics, Beijing Institute of Technology, 100081 Beijing, China}

\author{Chui-Zhen Chen}
\affiliation{School of Physical Science and Technology, Soochow University, Suzhou 215006, China}

\author{Hoi Chun Po\,\orcidlink{0000-0002-5773-0712}}\email{hcpo@ust.hk}
\affiliation{Department of Physics, Hong Kong University of Science and Technology, Clear Water Bay, Hong Kong, China}

\author{Jiangping Hu\,\orcidlink{0000-0002-4837-7742}}\email{jphu@iphy.ac.cn}
\affiliation{New Cornerstone Science Laboratory,  Institute of Physics, Chinese Academy of Sciences, Beijing 100190, China}

\author{Xianxin Wu\,\orcidlink{0009-0004-3916-7120}}\email{xxwu@itp.ac.cn}
\affiliation{CAS Key Laboratory of Theoretical Physics, Institute of Theoretical Physics, Chinese Academy of Sciences, Beijing 100190, China}

\begin{abstract}
The recent discovery of high-$T_{\rm c}$ superconductivity in bilayer nickelates has attracted enormous attention. Despite dominant theoretical studies pointing to interlayer pairing, the pairing symmetry is strongly debated due to conflicting evidence on nodeless bulk gaps and $d$-wave-like transport signatures. 
Here, we demonstrate that a nodeless interlayer pairing can naturally mimic $d$-wave Andreev signatures when the point contact is not evenly coupled to the two layers.
The phenomenon can be understood by recognizing how a contact potential, which is odd under the layer-exchanging mirror symmetry, acts like an effective Zeeman field on the layer pseudospin. This pseudo field then binds in-gap Andreev states that sweep through the gap with the width and strength of the barrier. We further demonstrate in a realistic model for La$_3$Ni$_2$O$_7$ that the simulated subgap conductance for a top-layer barrier can reproduce the orientation dependent ZBCP usually interpreted as a fingerprint of $d$-wave pairing. By revealing this mirror-selective Andreev response from an interlayer pairing state, our work offers a natural scenario to reconcile the conflicting experimental evidence in bilayer nickelates and advances our understanding of unconventional superconductivity.
\end{abstract}

\maketitle

\runinsec{Introduction}The pairing interaction of an unconventional superconductor is written in the sign structure of its superconducting gap, and reading that sign has been a central goal for decades~\cite{Sigrist1991,Monthoux1991,Scalapino2012,VanHarlingen1995,TsueiKirtley2000}. In the cuprates the $d$-wave order parameter changes sign under a real-space rotation. This property was the basis for the pioneering tricrystal experiments and decisive corner-junction half-flux-quantum tests of the pairing symmetry~\cite{Wollman1993,Tsuei1994}. At a $(110)$ surface it also binds a zero-energy Andreev bound state, seen as a zero-bias conductance peak (ZBCP)~\cite{Hu1994,TanakaKashiwaya1995,KashiwayaTanaka2000,Deutscher2005}. Identifying the sign-reversed pairing in the multi-orbital iron pnictides~\cite{Kamihara2008,Paglione2010,Stewart2011} is a harder problem. Their in-plane extended $s_\pm$-wave pairing is typically fully gapped and reverses sign between the hole and electron Fermi surfaces (FSs) that are well separated in momentum space. Unlike a $d$-wave gap, the sign-reversal is not connected by any real-space point-group transformation~\cite{Mazin2008,Kuroki2008,Hirschfeld2011}. This motivated a decade of proposals and experiments to read this relative sign change~\cite{Ghaemi2009,Christianson2008,Inosov2010,Hanaguri2010,ChenTsuei2010}. Nevertheless, an extended $s_{\pm}$ gap does not generically reproduce the surface phenomenology of a $d$-wave superconductor. Surface calculations for the pnictides find no zero-energy peak in the density of states~\cite{OnariTanaka2009}, and interface bound states appear only when the contact provides strong interpocket scattering, with their energies depending on sample-dependent parameters~\cite{Golubov2009,Sperstad2009,AraujoSacramento2009}.

The recently discovered bilayer nickelate La$_3$Ni$_2$O$_7$ crystallizing in a bilayer structure now raises the same question in the third high-temperature superconducting family, alongside the cuprates and the iron pnictides~\cite{Sun2023}. The bulk crystal superconducts near 80~K under pressure exceeding 14~GPa~\cite{Sun2023}, and the thin-film systems with compressive strain exhibit superconductivity at ambient pressure~\cite{Ko2025,Zhou2025Films}.
Within each bilayer, two NiO$_2$ layers connected by mirror symmetry are strongly coupled through the inner apical oxygen, rendering both $d_{x^2-y^2}$ and $d_{z^2}$ orbitals active~\cite{YaoDX,YZhang2023,Lechermann2023,Hirofumi2023possible,XWu,HHWen2023,XJZhou2023}.
Owing to the strong interlayer antiferromagnetic coupling, theoretical studies have proposed interlayer pairing as the dominant pairing component~\cite{Wang327prb,Hirofumi2023possible,XWu,ZhangGM_cpl2023,FangYang327prl,lu2024interlayer,zhang2024structural,HYZhangtype2,liao2023electron,WeiLi327prl,YifengYang327prb,Dagotto327prb,YifengYang327prb2,tian2023correlation,YiZhuangYouSMG,luo2023hightc,ryee2024quenched,ChenHH2025,zhan2024cooperation,PhysRevB.111.144514,CLe2025,GuanGJ2025,JZhan_QM}.
Because mirror reflection classifies the electronic states into mirror-even and mirror-odd sectors, this interlayer pairing naturally generates a symmetry-enforced $s_\pm$ state, placing opposite-sign gaps on the two mirror-sector FSs~\cite{ZhangQPI2025}.
Regarding the pairing symmetry, experimental measurements have revealed conflicting evidence in bulk and thin-film systems.
Point-contact Andreev-reflection spectroscopy~\cite{Andreev1964,Blonder1982} on pressurized single crystals reports a ZBCP, attributed to a $d$-wave component in a multi-gap fit~\cite{Liu2025SciChina} or to a nodal $d$-wave gap whose spectrum depends on the junction orientation~\cite{CaoPeng2025}. In contrast, another point-contact experiment reveals a two-gap $s$-wave spectrum and attributes its high-temperature ZBCP to a more transparent contact~\cite{Guo2025NatComm}. In thin-film systems, angle-resolved photoemission spectroscopy (ARPES)~\cite{Shen2026Science,Nie_NP2026} and scanning tunneling microscopy (STM)~\cite{Fan2025STM,XueQK_STM2026} find nodeless gaps, consistent with $s$-wave or interlayer $s_\pm$ pairing.
These puzzling experimental discrepancies present a theoretical challenge: reconciling a nodeless bulk with orientation-dependent $d$-wave-like transport~\cite{Yang2025Fano,Wang2025discriminate}. 

In this work, we explore the unique properties and Andreev reflection signatures associated with the proposed interlayer pairing and demonstrate how its interface scattering physics differs from conventional in-plane pairing.
We study the Andreev characteristics of interlayer pairing in a minimal bilayer model and apply these findings to bilayer nickelate. We demonstrate that interlayer pairing turns a mirror-odd contact potential into an effective Zeeman field, although nothing in the system carries a magnetic moment or breaks time-reversal symmetry. This effective Zeeman field on the layer pseudospin winds a mirror-mixing angle $\theta_{\rm M}$ with the width $L_{\rm B}$ and strength $V_{\rm o}$ of the barrier and binds in-gap Andreev states. These are unique features of the interlayer pairing. In the basis of mirror eigenstates, where the pairing sign is tied to the mirror eigenvalue, the same mirror-odd barrier mixes the two sectors and opens the sign-reversing Andreev channel. When applied to La$_3$Ni$_2$O$_7$, owing to its complementary Fermi surface geometry, a barrier with a mirror-odd component reproduces the three hallmarks usually read as $d$-wave: a ZBCP, strong orientation dependence and side dips. A fully gapped interlayer $s_\pm$ state thus mimics the Andreev signature of a nodal $d$-wave gap, and the mirror-selective Andreev reflection of the interlayer pairing offers a natural resolution to the conflicting experimental evidence in point contact measurements.

\begin{figure}[tbp!]
	\centering
	\includegraphics[width=1.0\columnwidth]{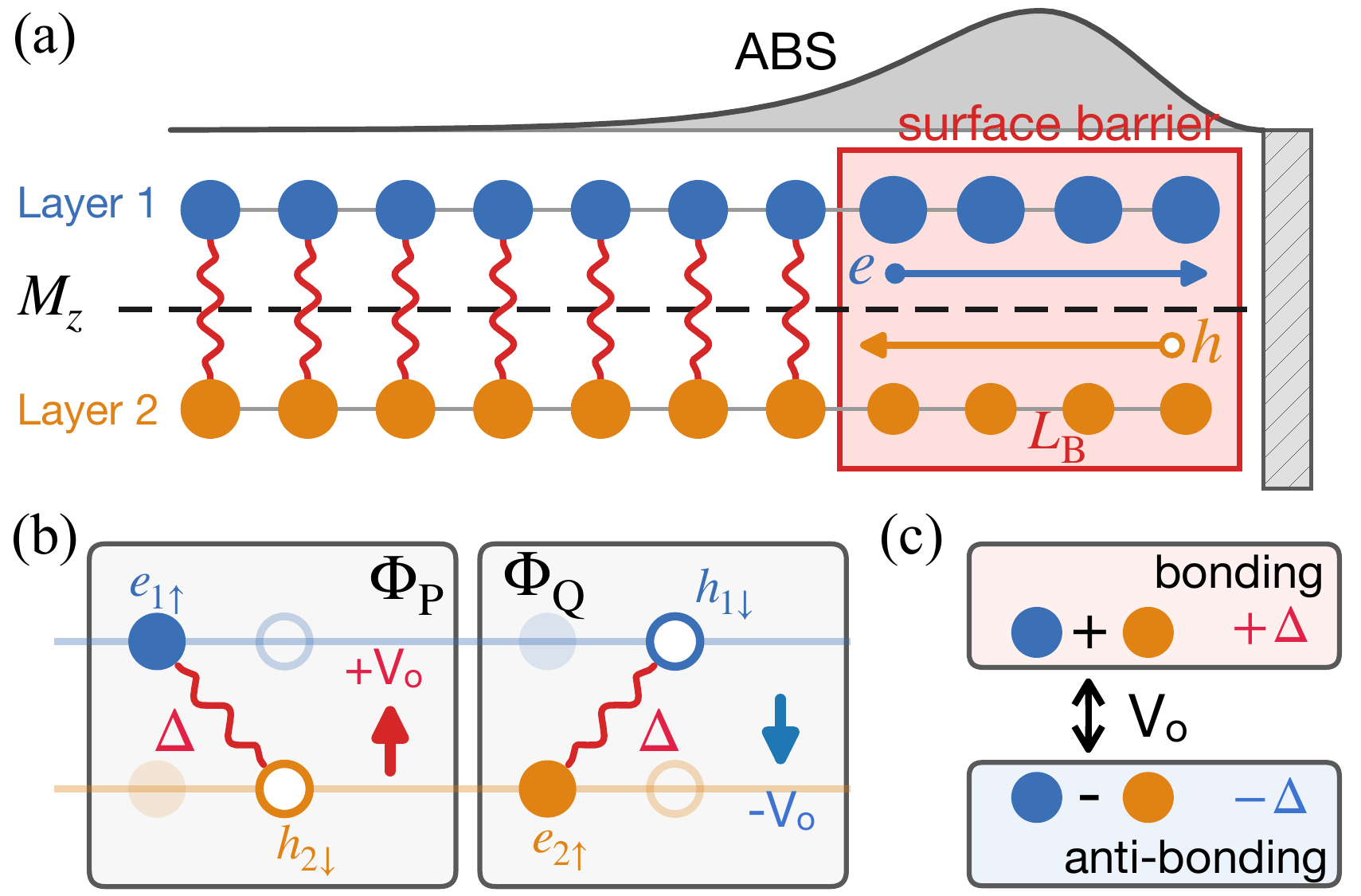}
	\caption{\label{fig1} (a) Sketch of the minimal bilayer model with a surface barrier on the last $L_{\rm B}$ sites and a hard wall. The wavy lines denote the interlayer pairing $\Delta$, the unequal circle sizes in the barrier denote the mirror-odd potential $\pm V_{\rm o}$ on the two layers, the arrows mark the electron ($e$) and the Andreev-reflected hole ($h$), and the shaded profile sketches the Andreev bound state (ABS). (b) The partner blocks $\Phi_{\rm P}$ and $\Phi_{\rm Q}$ in Eq.~(\ref{eq:blocks}), with the electron (filled circles) and the hole (open circles) of each block, the interlayer pairing $\Delta$ (wavy line) and the sign of $V_{\rm o}$ (arrows). (c) The two mirror sectors, with gaps $\pm\Delta$, coupled by $V_{\rm o}$.}
\end{figure}

\runinsec{ABSs from the mirror-odd barrier} We first study Andreev bound states (ABSs) for the interlayer pairing in a minimal bilayer model. We consider the bilayer model in contact with a potential barrier, where interlayer coupling and pairing are included, with a potential barrier on the last $L_{\rm B}$ sites, as sketched in Fig.~\ref{fig1}(a).
In the Nambu basis $\Psi=(c_{1k\uparrow},\,c_{2k\uparrow},\,c_{1-k\downarrow}^{\dagger},\,c_{2-k\downarrow}^{\dagger})^{T}$ the BdG Hamiltonian is
\begin{equation}
 H_{k} = \xi_{k}\ell_0\tau_z +t_{\perp}\ell_x\tau_z + \Delta\ell_x\tau_x + [V_{\rm o}(x)\ell_z+V_{\rm e}(x)\ell_0]\tau_z,
 \label{eq:toy}
\end{equation}
with $\xi_{k}=-2t\cos k-\mu$, $\ell$ and $\tau$ the layer and particle-hole Pauli matrices. $t_\perp$ and $\Delta$ are the interlayer coupling and pairing, generating symmetry-enforced $s_\pm$-wave pairing with sign-reversal pairing on the bonding and antibonding sectors, as sketched in Fig.~\ref{fig1}(c).
$V_{\rm o}(x)$ and $V_{\rm e}(x)$ are the mirror-odd and mirror-even components of the nonmagnetic contact potential.

To reveal the interesting feature of $V_{\rm o}$, we first neglect the interlayer coupling. As the pairing selects Andreev partners across the layers, we regroup the components into the two pairs $\Phi_{\rm P}=(c_{1k\uparrow},\,c_{2-k\downarrow}^{\dagger})^{T}$ and $\Phi_{\rm Q}=(c_{2k\uparrow},\,c_{1-k\downarrow}^{\dagger})^{T}$, shown in Fig.~\ref{fig1}(b). In this partner frame $H$ can be block diagonal, and each block is a plain $s$-wave problem,
\begin{equation}
 H_{\rm P/Q} = \xi_{k}\tau_z + \Delta\tau_x \pm V_{\rm o}(x)\tau_0 + V_{\rm e}(x)\tau_z .
 \label{eq:blocks}
\end{equation}
Two features of Eq.~(\ref{eq:blocks}) carry all the physics. First, although the gap in each block is $+\Delta$, the sign reversal has not disappeared; it has moved from the gap into the choice of partners, which pair an electron with a hole from the opposite layer. Second, the two barrier parities separate. The mirror-even part enters as $\tau_z$, the ordinary potential of BdG theory, with opposite signs for electron and hole. The mirror-odd part enters as $\tau_0$, acting with the same sign on the electron and the hole of a block. Therefore, with the layer acting as the pseudospin, the mirror-odd potential $V_{\rm o}$ enters the two blocks as an effective Zeeman field which is confined to the barrier region, and the $H_{\rm P/Q}$ of Eq.~(\ref{eq:blocks}) can be regarded as the Hamiltonian of a conventional superconductor with a Zeeman field confined to the barrier, although the system carries no magnetic moment. As $H_{\rm P/Q}$ differ only by the sign of $V_{\rm o}$ and are related by particle-hole symmetry, we work in $\Phi_{\rm P}$ below.

\begin{figure}[tbp]
	\centering
	\includegraphics[width=1.0\columnwidth]{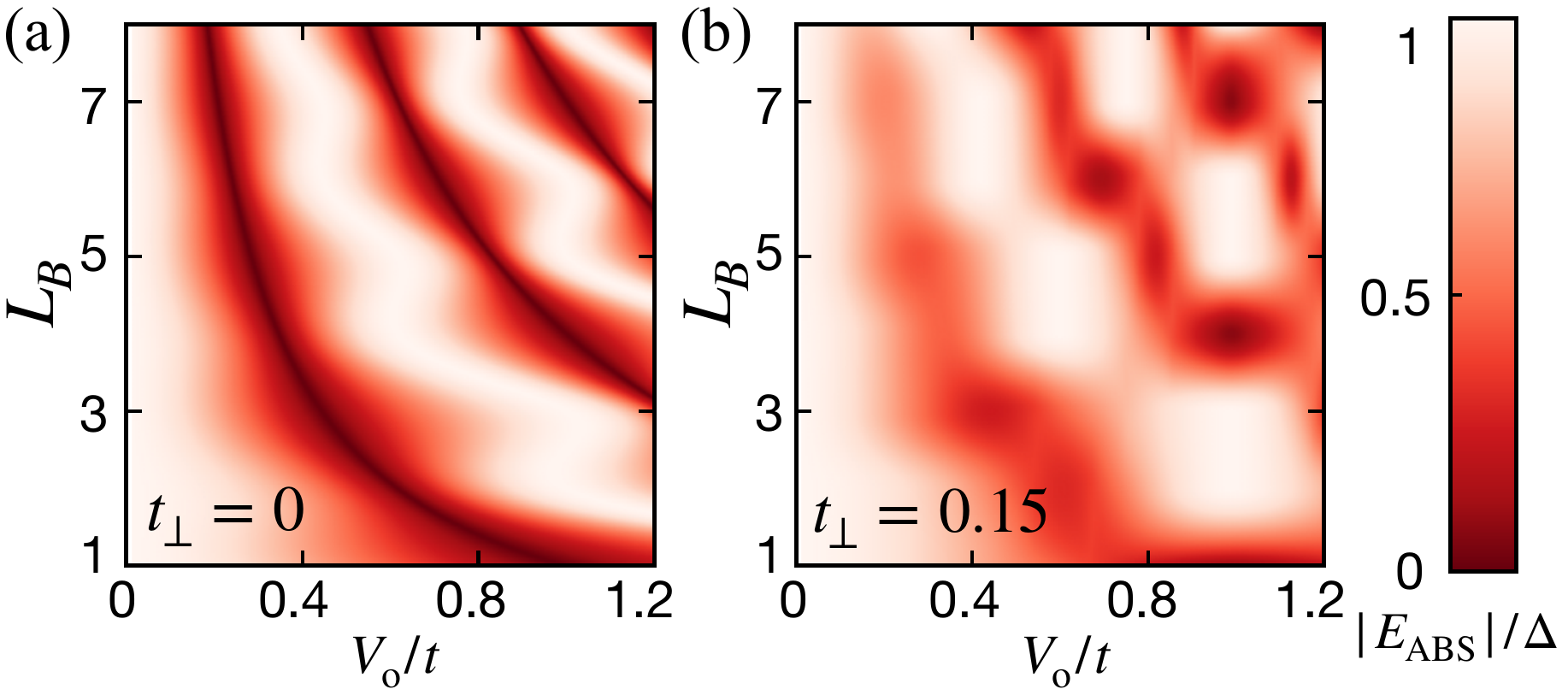}
	\caption{\label{figEABS} Andreev bound-state energy $|E_{\rm ABS}|/\Delta$ of the junction shown in Fig.~\ref{fig1}(a) with a mirror-odd barrier over the $(V_{\rm o},L_{\rm B})$ plane, at $\{\mu,\Delta\}=\{0,\,0.01t\}$ and (a) $t_\perp=0$, (b) $t_\perp=0.15t$. The energies are computed at integer $L_{\rm B}$ and interpolated in between.}
\end{figure}

Because $V_{\rm o}$ acts as a Zeeman field, we expect Andreev bound states around the interface, as at a ferromagnet--superconductor junction~\cite{Romeo2015}. Inside the barrier this effective Zeeman field $V_{\rm o}$ gives the electron and its opposite-layer Andreev-partner hole different Fermi momenta, $k_{\rm e}$ and $k_{\rm h}$, defined by $\xi(k_{\rm e})=E-V_{\rm o}$ and $\xi(k_{\rm h})=-(E-V_{\rm o})$. One round trip carries the electron across the barrier and back, an Andreev reflection into the hole, the hole across and back, and a second Andreev reflection. Electron and hole carry a U(1) phase and the round trip returns the phase difference $\theta_{\rm M}=2L_{\rm B}(k_{\rm h}-k_{\rm e})$, which is the counterpart of the spin-mixing angle of a magnetically active interface~\cite{Tokuyasu1988,Fogelstrom2000}. If we neglect the phase shift in scattering at the barrier boundary, which goes to zero for a barrier that turns on smoothly on the scale of the Fermi wavelength, the energy of in-gap Andreev bound states is given by
\begin{equation}
 E_{\rm ABS}=\pm\Delta\cos\frac{\theta_{\rm M}}{2},
 \label{eq:abs}
\end{equation}
with $\theta_{\rm M}\simeq2L_{\rm B}\Big[\arcsin\frac{V_{\rm o}-\mu}{2t}
      +\arcsin\frac{V_{\rm o}+\mu}{2t}\Big]$~\cite{SM}. In Fig.~\ref{figEABS}(a), the numerically computed $E_{\rm ABS}$ is shown as a function of $V_{\rm o}$ and $L_{\rm B}$ at $t_\perp=0$, where $E_{\rm ABS}$ can reach zero in several parameter regimes.

We now restore $t_\perp$ and two partner blocks $\mathrm{P}$ and $\mathrm{Q}$ are coupled. Stack the blocks of Eq.~(\ref{eq:blocks}) into $\bm{\Phi}=(\Phi_{\rm P},\Phi_{\rm Q})^{T}$ and let $\lambda$ be the Pauli matrices in $\bm{\Phi}$ space. $H_{\rm P/Q}$ in Eq.~(\ref{eq:blocks}) then reads $\xi\lambda_0\tau_z+\Delta\lambda_0\tau_x+V_{\rm o}\lambda_z\tau_0+V_{\rm e}\lambda_0\tau_z$ and is diagonal in $\lambda$, and interlayer coupling adds the only term that is not, $t_\perp\lambda_x\tau_z$. The electron and the hole of a partner pair therefore move in their own pseudospin fields $\mathbf{B}_{\rm e/h}=(\pm t_\perp,\,0,\,V_{\rm o})$, which share the magnitude $B=\sqrt{V_{\rm o}^{2}+t_\perp^{2}}$ but tilt to opposite sides of $\lambda_z$ by $\chi=\arctan(t_\perp/V_{\rm o})$. Crossing the barrier and returning, each of them turns the $\bm{\Phi}$-pseudospin through $e^{-\mathrm{i}\Theta\hat{\mathbf{B}}_{\rm e/h}\cdot\bm{\lambda}}$ with $\Theta\equiv2L_{\rm B}\arcsin(B/2t)$, an SU(2) rotation in place of the U(1) phase of the $t_\perp=0$ case. The large Fermi phase enters with opposite signs for the electron and the hole and cancels between them, which is the Andreev approximation~\cite{Andreev1964}, while the small phase $2L_{\rm B}E/v_{\rm F}$ enters with the same sign and is negligible for $L_{\rm B}\ll v_{\rm F}/\Delta$. One round trip is therefore a composition of two rotations about axes separated by $2\chi$, which is itself a rotation by $\Omega$ about a new axis $\hat{\mathbf{m}}$ tilted out of the plane of $\hat{\mathbf{B}}_{\rm e}$ and $\hat{\mathbf{B}}_{\rm h}$,
\begin{equation}
 E_{\rm ABS}=\pm\Delta\cos\frac{\Omega}{4},\
 \cos\frac{\Omega}{2}=\cos^{2}\Theta-\sin^{2}\Theta\cos2\chi .
 \label{eq:compose}
\end{equation}
The bound-state energy condition is Eq.~(\ref{eq:abs}) with $\theta_{\rm M}$ replaced by $\Omega/2$, and $E_{\rm ABS}$ is again its short-barrier limit~\cite{SM}. At $t_\perp=0$ the two axes align, $\cos(\Omega/2)=\cos2\Theta$, and Eq.~(\ref{eq:abs}) is recovered exactly, $\theta_{\rm M}=2\Theta$. In Fig.~\ref{figEABS}(b), the numerically computed $E_{\rm ABS}$ is shown as a function of $V_{\rm o}$ and $L_{\rm B}$ at $t_\perp=0.15t$, where the minimum $E_{\rm ABS}$ is close to zero.

The origin of Andreev bound states with finite $t_\perp$ can also be understood in the band basis with a sign-reversal pairing gap, similar to the well-known iron-based superconductors~\cite{Mazin2008,Kuroki2008,Ghaemi2009}.
Transforming to the bonding and antibonding basis, the Hamiltonian reads $\tilde{H}_{k} = [\xi_{k}\ell_0+t_{\perp}\ell_z]\tau_z + \Delta\ell_z\tau_x + V_{\rm o}(x)\ell_x\tau_z + V_{\rm e}(x)\ell_0\tau_z$.
It is apparent that the pairing is sign-reversed between two mirror sectors, a mirror-enforced $s_\pm$-wave pairing. Now $V_{\rm e}$ is diagonal and scatters states within each sector, while $V_{\rm o}$ is off-diagonal and mixes mirror-even and mirror-odd states. Within the barrier, the $V_{\rm o}$ potential induces an SU(2) rotation in two sectors as the electron and hole traverse the region. As the pairing term $\Delta\ell_z\tau_x$ is sign-reversed between the sectors, inter-sector mixing induces a $\pi$-phase shift in two consecutive Andreev reflections if they occur in opposite sectors. Tracking a quasiparticle through one round trip, the total scattering matrix is $e^{-2\mathrm{i}\eta}\ell_z\hat{\tilde U}_{\rm h}\ell_z\hat{\tilde U}_{\rm e}$, with $\eta=\arccos(E/\Delta)$ being the Andreev reflection phase~\cite{SM}. Solving the resonance condition for this matrix gives the in-gap ABS energy in Eq.~(\ref{eq:compose}).

\begin{figure}[bp]
	\centering
	\includegraphics[width=1.0\columnwidth]{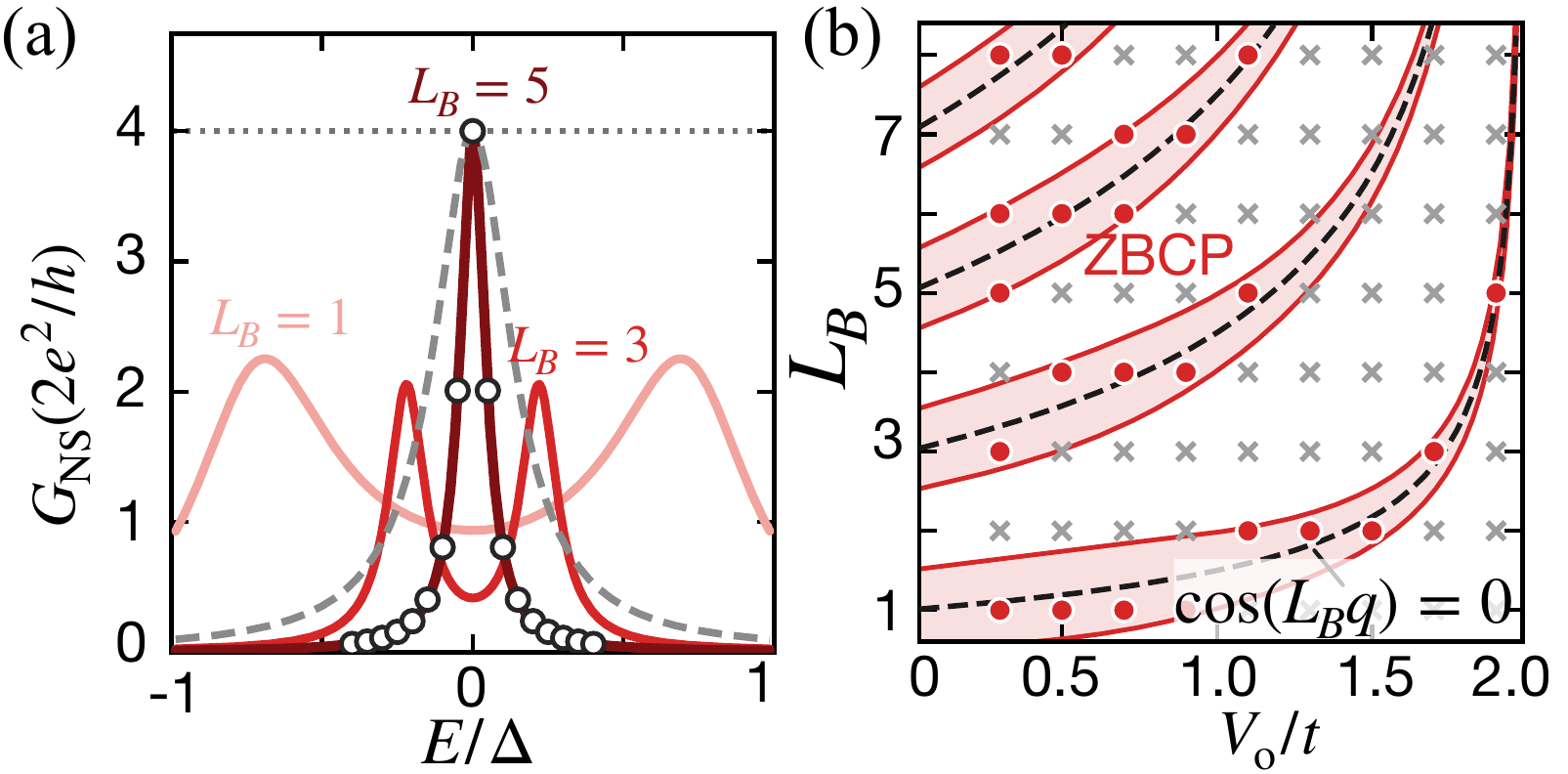}
	\caption{\label{figtoyG} Sub-gap conductance of the normal-metal--barrier--superconductor junction at $t_\perp=\mu=0$ and $\Delta=0.002t$. (a) $G_{\rm NS}(E)$ at $V_{\rm o}=2t\cos(\pi/10)$ for $L_{\rm B}=1$, $3$ and $5$, from light to dark. The dotted line is the unitarity bound $8e^{2}/h$ and the circles are the Lorentzian form for $L_{\rm B}=5$. The grey dashed line is the conductance at $t_\perp=0.5t$, $V_{\rm o}=1.658t$ and $L_{\rm B}=3$. (b) The ZBCP region (red) in the $(V_{\rm o},L_{\rm B})$ plane, shaded and bounded by the red lines. The black dashed lines are the resonances $\cos(L_{\rm B}q)=0$.}
\end{figure}

\runinsec{Differential conductance of the minimal junction} We further consider a metal--$V_{\rm o}$-barrier--superconductor junction within the same minimal model. The junction consists of a normal lead, a mirror-odd barrier of width $L_{\rm B}$, and a superconductor with interlayer pairing. Coupling the in-gap Andreev states discussed above to the continuum of the normal lead turns them into leaky Andreev states, which appear as resonances in the differential conductance.
Below the superconducting gap the superconductor has no state to transmit into, so an incident electron either returns as an electron or is Andreev reflected as a hole. The conductance through the Andreev process is $G_{\rm NS}=(4e^{2}/h)\,\mathrm{Tr}\,r_{\rm he}^{\dagger}r_{\rm he}$, with $r_{\rm he}$ the electron-to-hole reflection matrix~\cite{Blonder1982,Beenakker1997}.
We first analyze the simplified case at $t_\perp=0$ and $\mu=0$. Within the Blonder-Tinkham-Klapwijk (BTK) formalism, the zero-bias conductance reads
\begin{equation}
 G_{\rm NS}(0)=\frac{8e^{2}}{h}
 \frac{1-\nu^{2}}{1-\nu^{2}\cos^{2}(2L_{\rm B}q)},
 \label{eq:toyG}
\end{equation}
where $\nu\equiv\cos q=V_{\rm o}/2t$, and $q$ and $\pi-q$ are the wave vectors of particle and hole within the barrier~\cite{SM}. These conductance resonances originate from the coherent addition of multiple electron-hole round trips inside the barrier region. After each round trip, the quasiparticle can escape into the normal lead as an Andreev-reflected hole. When the total accumulated phase through propagation, normal-reflection, and Andreev processes during a round trip is an integer multiple of $2\pi$, the hole amplitudes emitted after different numbers of round trips are in phase and interfere constructively. Andreev reflection is then strongly enhanced and normal reflection is suppressed, producing a conductance resonance. In the present limit, $4L_{\rm B}q$ is the relative propagation phase accumulated by the electron and hole during one complete round trip. The conductance reaches its maximal value at $2L_{\rm B}q=m\pi$ with normal-reflection and Andreev phases adding to $2\pi$. An even $m$ means perfectly transmitting, thus the resonance condition is $V_{\rm o}=2t\cos[(n+\tfrac12)\pi/L_{\rm B}]$, where the phase-matched electron-hole orbit forms a leaky zero-energy Andreev state. Near the zero-energy resonance, the bias dependence takes the Lorentzian form $G_{\rm NS}(E)=\frac{8e^{2}}{h}\frac{(1-\nu^{2})^{2}}{(1-\nu^{2})^{2}+4\nu^{2}(E/\Delta)^{2}}$, with half width at half maximum $\Delta(1-\nu^{2})/2\nu$. At $E=0$, the resonant state mediates perfect Andreev reflection in both transport channels, giving the maximal conductance $G_{\rm NS}(0)=8e^2/h$.

 The resonance is the leaky counterpart of the bound state of Eq.~(\ref{eq:abs}). From the point of view of the effective Zeeman field, it is the resonant proximity effect of a ferromagnet--superconductor junction~\cite{Romeo2015}, and in the band basis it is the $\pi$ phase collected by two Andreev reflections in sectors of opposite gap sign, as at the $(110)$ surface of a $d$-wave superconductor. In Fig.~\ref{figtoyG}(a), the energy-dependent conductance is shown for several values of $L_{\rm B}$. The finite-energy sub-gap peaks for $L_{\rm B}=1$ and $3$ come from leaky in-gap Andreev states, and at $L_{\rm B}=5$ the state reaches zero energy and gives a sharp ZBCP. In Fig.~\ref{figtoyG}(b), the region of the $(V_{\rm o},L_{\rm B})$ plane in which a ZBCP occurs is shown, with boundaries derived in~\cite{SM}, and each resonance line, $\cos(L_{\rm B}q)=0$, runs near the center of a finite window over which the peak stays visible.

 Finite $\mu$ and $t_\perp$ detune the resonance. The two sectors acquire different Fermi momenta and the interlayer coupling weakens their mixing by the barrier, so the leaky Andreev states shift away from zero energy, although an exact zero-energy resonance can be recovered by retuning $V_{\rm o}$ and $L_{\rm B}$. When the shift is smaller than the resonance linewidth, the resonances at $\pm E$ overlap into a broad ZBCP, as the dashed curve in Fig.~\ref{figtoyG}(a) shows, and for larger detuning the peak moves off zero or splits into two finite-bias peaks. The mirror-odd barrier thus produces a ZBCP through resonant Andreev reflection in a superconductor with interlayer pairing, but the mirror-even barrier leaves the conductance fully gapped.

\runinsec{Application to bilayer nickelates La$_3$Ni$_2$O$_7$}
After studying the bilayer toy model, we further explore Andreev bound states in the high-$T_{\rm c}$ bilayer nickelate with a realistic model. For its normal state we use the two-orbital tight-binding model of Ref.~\cite{CLe2025}~\cite{SM}.
The apical-oxygen $\sigma$-bond between the $d_{z^2}$ orbitals splits each band into bonding and antibonding combinations, the two sectors of $\mathcal{M}_z=\ell_x\sigma_0\tau_0$. Here $\sigma$ acts on the orbitals. Driven by the strong interlayer antiferromagnetic coupling, the dominant pairing is expected to emerge in the interlayer channel.
Thus we take $\Delta_{\rm pair}=(\Delta_{\perp}\,\ell_x+\Delta_{\parallel}\ell_0)\sigma_0\tau_x$, with $\Delta_{\perp}$ and $\Delta_{\parallel}$ being the interlayer and intralayer pairing amplitudes~\cite{Wang327prb,XWu,zhan2024cooperation,JZhan_QM}. 
The amplitudes are chosen to reproduce the multiple gap structure observed in ARPES and STM measurements~\cite{Shen2026Science,Nie_NP2026,Fan2025STM,XueQK_STM2026,Guo2025NatComm}.

\begin{figure}[!t]
	\centering
	\includegraphics[width=0.92\columnwidth]{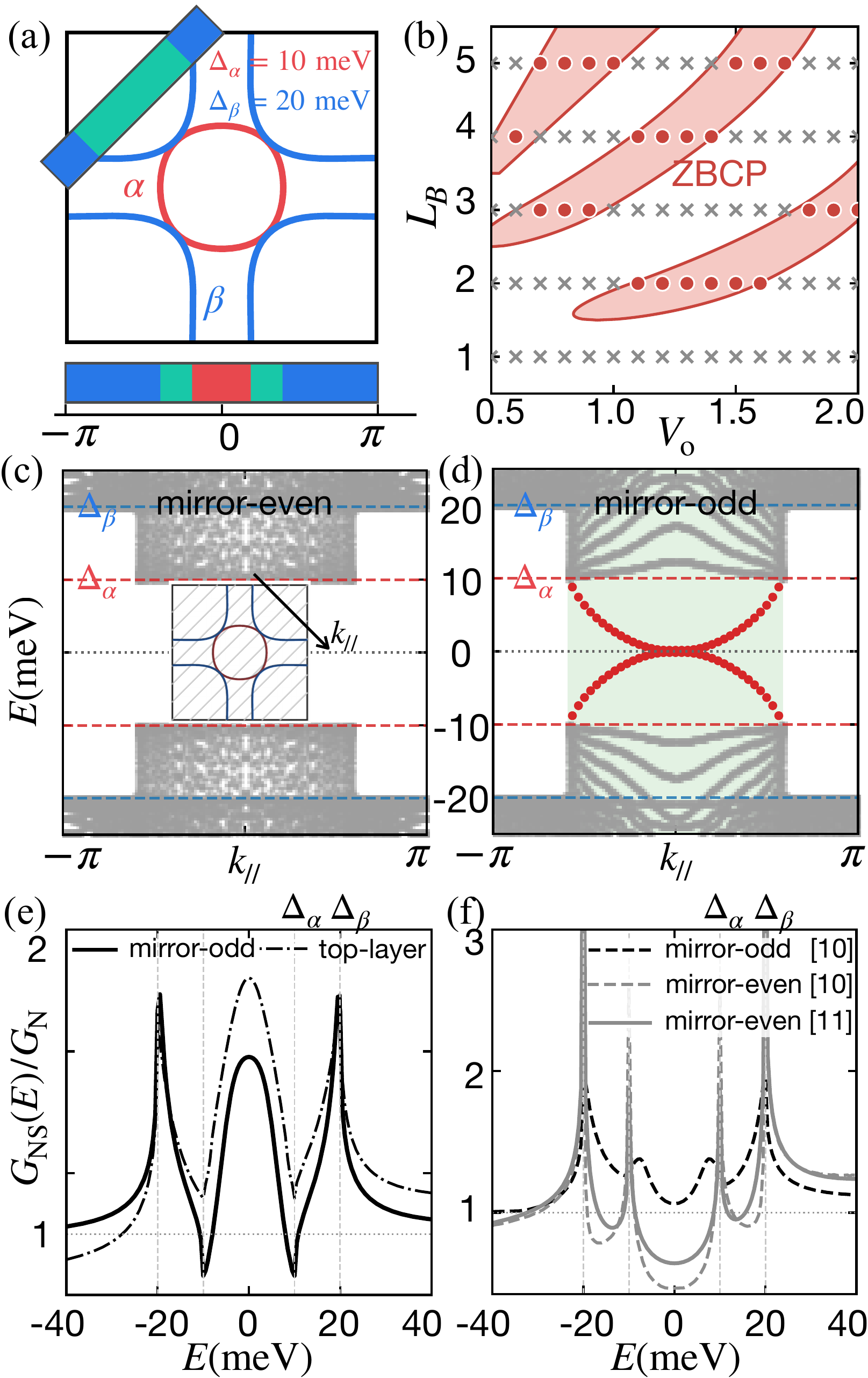}
\caption{\label{fig327} Coexistence channels, surface spectra and point-contact conductance of La$_3$Ni$_2$O$_7$ at $\mu=0$ with $\Delta_\alpha=10$~meV and $\Delta_\beta=20$~meV. (a) The bonding ($\alpha$, red) and antibonding ($\beta$, blue) Fermi surfaces and their projections onto $k_\parallel$ along $[11]$ and $[10]$, green where both bands cross the Fermi level. (b) In red, the $(V_{\rm o},L_{\rm B})$ region in which the conductance along $[11]$ with a mirror-odd barrier shows a ZBCP. (c),(d) BdG spectrum of a slab cut along $[11]$ for a mirror-even ($V_{\rm e}=1.0$) and a mirror-odd ($V_{\rm o}=1.0$) barrier of width $L_{\rm B}=1$, with the boundary-localized in-gap branch in red, the continuum in grey and the coexistence window shaded green. The inset of (c) shows the Fermi surfaces and the direction of $k_\parallel$. (e),(f) $G_{\rm NS}(E)$ normalized to the zero-bias normal-state conductance $G_{\rm N}$. (e) Along $[11]$, the mirror-odd barrier with $V_{\rm o}=0.70$, $L_{\rm B}=3$ (solid) and a top-layer barrier of $0.70$ on one site (dash-dotted). (f) The mirror-odd barrier along $[10]$ at the same parameters (black dashed) and the mirror-even barrier with $V_{\rm e}=0.65$, $L_{\rm B}=3$ along $[11]$ (grey solid) and $[10]$ (grey dashed).}
\end{figure}

The two FSs of the bilayer nickelate have complementary shapes, the bonding pocket closing around $\Gamma$ and the antibonding sheet enclosing the zone corner, as shown in Fig.~\ref{fig327}(a). Along the diagonal direction, two orbitals are decoupled and Fermi surfaces of different mirror sectors are of $d_{x^2-y^2}$ character with close Fermi momenta, resembling the toy model in Eq.~(\ref{eq:toy}) with negligible $t_\perp$. Transport through the cross-sector channel needs both sectors to propagate. When the bilayer nickelate is in contact with a mirror-odd barrier $V_{\rm o}$ along a certain direction, the transverse momentum $k_\parallel$ parallel to the interface is conserved. Scattering by the mirror-odd barrier allows successive Andreev reflections to occur in opposite mirror sectors and hence experience opposite pairing signs. This process requires that both sectors cross the Fermi level at that $k_\parallel$. We call such momenta coexistence channels. Their fraction is fixed by the FS geometry and depends strongly on the contact direction, $59\%$ of the zone along $[11]$ and $20\%$ along $[10]$, as shown in Fig.~\ref{fig327}(a). This kinematic fraction determines how much of the zone can carry the sub-gap Andreev states, while the per-channel line shape is set by $\theta_{\rm M}$ in the scattering.
An open-slab calculation of bilayer nickelates confirms the above analysis in the toy model, as shown in Fig.~\ref{fig327}(c),(d). A mirror-odd surface barrier binds a branch of sub-gap Andreev states that disperses with $k_\parallel$ and approaches zero energy inside the coexistence window. The energies of these in-gap Andreev states depend on the width and strength of the barrier and generally do reach zero. In contrast, a mirror-even barrier and a bare surface leave the spectrum fully gapped.

We further study the differential conductance for junctions with mirror-even and mirror-odd barriers in bilayer nickelates. In experiments, a point contact measures the differential conductance, set at zero temperature by the interface scattering matrix at the bias energy~\cite{TYChen2008,Daghero2011}. For a normal-metal-superconductor contact,
\begin{equation}
G_{\rm NS}(E)=\frac{2e^2}{h}\sum_{k_\parallel}\Big[N_{\rm e}-\mathrm{Tr}\,r_{\rm ee}^\dagger r_{\rm ee}+\mathrm{Tr}\,r_{\rm he}^\dagger r_{\rm he}\Big]_{k_\parallel},
 \label{eq:gns}
\end{equation}
the multichannel BTK form~\cite{Blonder1982,TakaneEbisawa1992,Beenakker1997}, with $r_{\rm ee}$ and $r_{\rm he}$ the normal and Andreev reflection blocks and $N_{\rm e}$ the number of incoming electron channels. We evaluate Eq.~(\ref{eq:gns}) for a junction made of a normal lead with the same bands, a barrier covering $L_{\rm B}$ interface sites, and bilayer nickelates~\cite{SM}.
In Figs.~\ref{fig327}(e) and \ref{fig327}(f), the conductance along the $[11]$ and $[10]$ directions is shown for both mirror-odd and mirror-even barriers.
For a pure mirror-odd barrier, the conductance shows a broad ZBCP along $[11]$ and a zero-bias dip along $[10]$, as shown in Fig.~\ref{fig327}(e),(f). Each coexistence channel resembles the minimal junction of Fig.~\ref{figtoyG}, with a resonance condition set by its own Fermi momenta. The channels close to resonance carry a zero-bias peak as in Fig.~\ref{figtoyG}(a), and the others peak at finite bias~\cite{SM}. The ZBCP is broad because the energy of the Andreev bound states disperses with $k_\parallel$ through $\theta_{\rm M}$, and the zone average retains a zero-centered enhancement across the coexistence window. The conductance is an open-junction observable, so this broad peak is the integrated weight of various momentum-resolved resonances and does not require any single level to sit exactly at zero. The strong orientation dependence is related to complementary FS geometries and different coexistent fractions.
In the mirror basis a mirror-even barrier is block diagonal, so the conductance recovers the conventional $s$-wave BTK line shape along both directions, as shown in Fig.~\ref{fig327}(f). Moreover, the peak along $[11]$ is accompanied by side dips below the normal-state conductance, as shown in Fig.~\ref{fig327}(e). These qualitative features remain unchanged even when the bonding $\gamma$ pocket appears~\cite{SM}.
The conductance profile depends on $V_{\rm o}$ and $L_{\rm B}$ parameters, but the broad ZBCP along $[11]$ direction survives across a wide parameter regime, as highlighted by red regions of Fig.~\ref{fig327}(b). Importantly, this ZBCP is not restricted to pure mirror-odd potentials. For example, a barrier potential confined to the top layer, an equal mix of odd and even components, generates a ZBCP, as shown in Fig.~\ref{fig327}(e).

\runinsec{Summary and discussion}
Our study highlights two unique properties of interlayer pairing. First, the layer degree of freedom acts as a pseudospin, allowing a nonmagnetic mirror-odd potential to behave as an effective pseudospin Zeeman field. This pseudospin coupling gives rise to a mirror-selective Andreev response for barriers of different mirror parity and, when sufficiently strong, can create a mismatch between the paired layer-pseudospin sectors that favors finite-momentum, FFLO-like interlayer pairing~\cite{OhZhang2025PDW,FanVishwanath2025}. Second, interlayer pairing produces symmetry-enforced, opposite-sign gaps in the two mirror sectors, which can reside on closely spaced Fermi surfaces derived from the same underlying electronic states. This differs from the canonical \(s_{\pm}\) state driven by in-plane pairing in iron-based superconductors, where the sign reversal typically occurs between Fermi pockets that are widely separated in momentum space and often have different orbital character. Coupling between the closely spaced mirror sectors can occur with little momentum transfer and is governed by the mirror parity of the perturbation. By contrast, scattering between the opposite-sign pockets in iron-based superconductors lacks such a mirror-selection rule, requires a large momentum transfer, and strongly depends on the orbital structure associated with pockets. These constraints limit iron-based systems to finite-energy in-gap conductance peaks, rendering a robust ZBCP difficult to realize~\cite{Golubov2009,Sperstad2009,Burmistrova2013}.

Our calculations of the differential conductance for mirror-even and mirror-odd barriers in bilayer nickelates offer a reconciliation of conflicting point-contact measurements. In measurements without controlled junction orientation, one study reported a two-gap $s$-wave-like conductance spectrum~\cite{Guo2025NatComm}, while another observed a ZBCP that was interpreted as evidence for a $d$-wave component~\cite{Liu2025SciChina}. Directional point-contact measurements have further revealed noticeable orientation dependence, including ZBCPs and side dips, and were thus interpreted as supporting predominantly $d$-wave pairing~\cite{CaoPeng2025}.
Our model calculations with interlayer pairing provide an alternative, consistent explanation for these discrepancies. For a mirror-odd barrier, the calculated differential conductance exhibits a strong in-plane orientation dependence and can develop a ZBCP along the $[11]$ direction and side dips owing to the complementary geometry of FSs. By contrast, a mirror-even barrier yields a two-gap $s$-wave conductance spectrum. Uncontrolled variations in the barrier parity and junction orientation may therefore account for differences among existing experiments. In particular, a ZBCP in point-contact spectroscopy is not, by itself, definitive evidence for $d$-wave pairing: interlayer pairing combined with a mirror-odd barrier can produce a qualitatively similar feature through resonant Andreev reflection.
This contact dependence provides a testable prediction: the $d$-wave-like ZBCP should be tunable through the layer asymmetry of the contact and disappear in the mirror-even limit of equal coupling to the two layers.
Finally, the fully gapped bulk spectrum associated with interlayer pairing is consistent with the nodeless superconducting gaps observed by ARPES and STM in ambient-pressure thin films~\cite{Shen2026Science,Fan2025STM}.
Taken together, these results show that interlayer pairing can reconcile a nodeless bulk gap with strongly contact-dependent and orientation-dependent conductance spectra in high-$T_{\rm c}$ bilayer nickelates.

\begin{acknowledgments}
Z.Z. and H.C.P. acknowledge support from the Future Science Awards Foundation and the Hong Kong Research Grants Council through grants No. 16310625 and AoE/P-604/25R.
S.Q. is supported by the National Natural Science Foundation of China (Grant No. NSFC-12304163).
C.-Z.C. acknowledges support from the National Key R\&D Program of China Grant (2022YFA1403700) and the Natural Science Foundation of Jiangsu Province Grant (BK20230066).
J.H. is supported by the National Natural Science Foundation of China (Grant No. NSFC-12494594) and the New cornerstone investigator program.
X.W. is supported by the National Key R\&D Program of China (Grant No.~2023YFA1407300) and the National Natural Science Foundation of China (Grants No.~12574151, 12447103 and 12447101).
\end{acknowledgments}

\bibliography{refs}

\clearpage

\end{document}